\documentclass[%
 reprint,
 amsmath,amssymb,
 aps,
]{revtex4-1}

\usepackage{graphicx}
\usepackage{dcolumn}
\usepackage{bm}

\begin{document}


\title{Study of charged particle multiplicity, average transverse momentum and azimuthal anisotropy in Xe+Xe collisions at $\sqrt{s_{NN}}$ = 5.44 TeV using AMPT model}
\author{Sourav Kundu}
\email{souravkundu692@niser.ac.in}
\author{Dukhishyam Mallick}
\email{dukhishyam.mallick@niser.ac.in}
\author{Bedangadas Mohanty}
\email{bedanga@niser.ac.in }
\affiliation{School of Physical Sciences, National Institute of Science Education and Research, HBNI, Jatni 752050, India}

\date{\today}
\begin{abstract}
We have studied the average charged particle density ($<$dN$_{ch}$/d$\eta$$>$), transverse momentum ($p_{\mathrm{T}}$) spectra, $<$$p_{\mathrm{T}}$$>$ and azimuthal anisotropies of inclusive charged particles produced in  Xe+Xe collisions at $\sqrt{s_{NN}}$ = 5.44 TeV using A Multiphase Transport Model (AMPT), which includes the deformation of Xe$^{129}$ nucleus. Calculations have been performed with the string melting version of AMPT model and compared with the recent measurements from the ALICE experiment. 
The model results over predict the measured $<$dN$_{ch}$/d$\eta$$>$ for central collisions, agree with the data for mid-central collisions and under predict the measurements for peripheral collisions. The centrality dependence of $<$$p_{\mathrm{T}}$$>$ of charged particles measured in ALICE is not reproduced by the model results. The calculated elliptic flow ($v_{2}$) from AMPT model overpredicts the ALICE measurements in central collisions but are consistent with the data in mid central collisions. We find that the model shows a mild centrality dependence of triangular flow and overestimates the ALICE measurements. Within the model framework,  we have also studied various collision configurations of Xe nuclei such as body-body, tip-tip, side-side and random. We find a strong dependence of the above observable on the collision configurations.
\end{abstract}
\maketitle
\section{\label{sec:level1}Introduction}
After a successful heavy-ion program at the Large Hadron Collider (LHC) and Relativistic Heavy-ion Collider (RHIC) facilities, observations such as large $v_{2}$ and its number of constituent quark scaling~\cite{v2}, jet quenching~\cite{jet}, suppression in the production of high $p_{\mathrm{T}}$ ($p_{\mathrm{T}}$ $>$ 8 GeV/$c$) hadrons compare to p+p collisions have confirmed the presence of a deconfined state of quarks and gluons (QGP)~\cite{rhic,alice} in ultrarelativistic heavy-ion collisions. In the ALICE experiment so far heavy-ion collisions have been carried out by Pb$^{208}$ nucleus~\cite{Pb}. Recently on 9$^{th}$ October 2017, ALICE collected data for Xe+Xe collisions~\cite{xe} over a time period of 8 hours. Xe nucleus has a moderate prolate deformation~\cite{deform} compared to spherical Pb nucleus. A deformed Xe nucleus collision allow us to probe a different initial condition compared to the collision of spherical Pb nucleus.\\
Deformed shape of Xe nucleus provide us with various type of collision configurations such as body-body, tip-tip and side-side depending upon the angle of colliding Xe nucleus with respect to reaction plane. Central heavy-ion collisions with a spherically symmetric nucleus such as Au, Pb always give a circular overlapping region in xy plane but in the case of deformed nucleus the overlapping region need not be circular. Measurement of azimuthal anisotropies and their fluctuations in various type of collision configurations and comparison with the results from the collision of the spherical nucleus will provide us the necessary input to constraining the initial condition models. Hydrodynamical model simulation predicts an increase of elliptic flow ($v_{2}$) by 10$\%$ for a deformed shape of Xe nucleus compared to the spherical shaped Xe nucleus in the central collisions and the effect of deformation on $v_{2}$ vanishes beyond the collision centrality class of higher than 15$\%$~\cite{hydro}.\\
Furthermore, heavy-ion collisions with deformed nucleus provide us with a better handle of background measurement for chiral magnetic effect (CME)~\cite{cme1,cme2}. Magnetic field produced due to the presence of spectator nucleons is a source of CME observable whereas $v_{2}$ acts as a background for CME observable. In the central collisions  both $v_{2}$ and magnetic field are less whereas in mid-central collisions both of them are large. So, the most preferable collisions to observe CME are those where the magnetic field is finite and $v_{2}$ is minimum and this type of scenario can be achieved in the collision of the deformed nucleus by selecting body-tip type collision configuration.\\ 
In a recent hydrodynamic based model calculations~\cite{hydro} of various observables in Xe+Xe collisions, the information of the deformed Xe nucleus has been considered. In this work, we present result from a transport based model study of Xe+Xe collisions after including the deformation information. We also compare the model results to those measured in the ALICE experiment~\cite{alicespectra,xe}. In addition, we also present results for different Xe+Xe collision configurations for observables such as $<$dN$_{ch}$/d$\eta$$>$, $p_{\mathrm{T}}$ spectra, $<$$p_{\mathrm{T}}$$>$ and azimuthal anisotropies.\\
This paper is organized as follows, in the next section we describe AMPT model and implementation of Xe+Xe collisions in AMPT~\cite{ampt1,ampt2} model, along with the specific configurations of Xe+Xe collisions. In section III we compare the AMPT model results with ALICE measurements and present our results on $<$dN$_{ch}$/d$\eta$$>$, $p_{\mathrm{T}}$ spectra, $<$$p_{\mathrm{T}}$$>$ and azimuthal anisotropies for different collision configurations. Finally in section IV we summarize our observations.\\
\section{\label{sec:level2}Implementation of Xe+Xe collisions in AMPT model}
We have used the AMPT model with string melting (SM) version. It is a hybrid transport model. This model contains four basic stages:\\
1)Initial condition: Based on the initial condition from HIJING~\cite{hijing}.\\
2)Partonic scattering: Zhang's parton cascade model~\cite{zhang} has been used for scattering among the partons. Parton scattering is calculated by using two body scattering cross section from pQCD with screening masses.\\
3)Hadronization: In SM version of AMPT soft partons are created from string and a quark coalescence model is used to combine parton into hadrons.\\
4)Hadronic interaction:  A Relativistic Transport (ART)~\cite{art} model is used for the evaluation of hadronic matter, which includes the interaction between hadrons.\\
For our study, we have used AMPT version 2.25t7 with partonic cross section of 3mb. String melting parameters $a$ and $b$ are 0.3 and 0.15 respectively. We have implemented the deformation information of Xe nucleus in AMPT by using a deformed Woods-Saxon~\cite{wood} profile,
\begin{equation}
\rho=\frac{\rho_{0}}{1+exp(|r-R|/d)}  
\end{equation}
\begin{equation}
R=R_{0}[1+\beta_{2}Y_{2}^{0}(\theta)+\beta_{4}Y_{4}^{0}(\theta)]
\end{equation}
$\rho_{0}$ is the normal nuclear density, $R_{0}$ is the radius of Xe nucleus, $d$ is the diffuseness parameter and $\beta_{2}$, $\beta_{4}$ are the deformed parameters for Xe nucleus. We have used $R_{0}$ = 5.4 fm, $d$ = 0.59 fm, $\beta_{2}$ = 0.162 and $\beta_{4}$ = -0.003~\cite{deform}. $Y_{l}^{m}(\theta)$ is the spherical harmonics. This implementation is similar to those for U+U collisions as discussed in~\cite{uurihan}\\
In this work, we have studied 3 type of collision configuration of Xe nucleus which are chosen by the orientation of semi-major axis and semi-minor axis of the colliding deformed Xe nucleus. We compare results from these configurations with the Xe+Xe collision which have a random orientation of deformed Xe nucleus (usual experimental situation). In addition, we also present the result of Xe+Xe collision where the Xe nucleus is considered to be spherically symmetric in order to see the effect of deformation on the observables studied. This configuration is created by setting the deformation parameters $\beta_{2}$ and $\beta_{4}$ to zero. Three type of special configurations which we have chosen for this analysis are named as body-body, tip-tip and side-side. Orientation unbiased configuration is named as random and collisions with spherical Xe nucleus is named as spherical.\\
Details about the angular orientation of various configurations in terms of $\theta$ and $\phi$ are given in Table~\ref{table:angle}.
\begin{table}
\caption{Details of angular configuration in Xe+Xe collisions. $t$ and $p$ in subscript denotes the target and projectile respectively.}
\label{table:angle}
\begin{center}
\begin{tabular}{|c|c|c|c|c|c|}
\hline
Configuration&$\theta_{p}$&$\theta_{t}$&$\phi_{p}$&$\phi_{t}$&Impact parameter direction\\
\hline
general&0-$\pi$&0-$\pi$&0-2$\pi$&0-2$\pi$&random\\
\hline
tip-tip&0&0&0-2$\pi$&0-2$\pi$&minor axis\\
\hline
body-body&$\pi$/2&$\pi$/2&0&0&major axis\\
\hline
side-side&$\pi$/2&$\pi$/2&$\pi$/2&$\pi$/2&minor axis\\
\hline
\end{tabular}
\end{center}
\end{table}
In simulation of random configuration both the colliding Xe nucleus are randomly rotated along the polar direction by using a uniform distribution in $\theta$ with a weight of sin$\theta$ and in an azimuthal direction with a uniform $\phi$ distribution. We have analyzed 50000 minimum bias events for each configuration for the analysis carried out in the paper.\\
\section{\label{sec:level3}Results}
In order to compare the model result with the ALICE experiment, we select those events which have at least 1 charged particle present in $|\eta|$ $<$ 1. This condition is similar to INEL$>$0 trigger used in the ALICE experiment~\cite{alicespectra,xe}. Centrality selection is done by bining in multiplicity distribution of charged particle. For centrality selection, we take the multiplicity distribution of charged particle in the $\eta$ range 2.8 $<$ $\eta$ $<$ 5.1 and -3.7 $<$ $\eta$ $<$ -1.7. This centrality selection, is chosen to mimic the centrality selection used in the ALICE experiment.\\
\subsubsection{\label{sec:level3}Multiplicity, Pseudorapidity distribution, Transverse momentum spectra and Mean Transverse momentum}
\begin{figure}[hbtp]
\centering 
\includegraphics[scale=0.4]{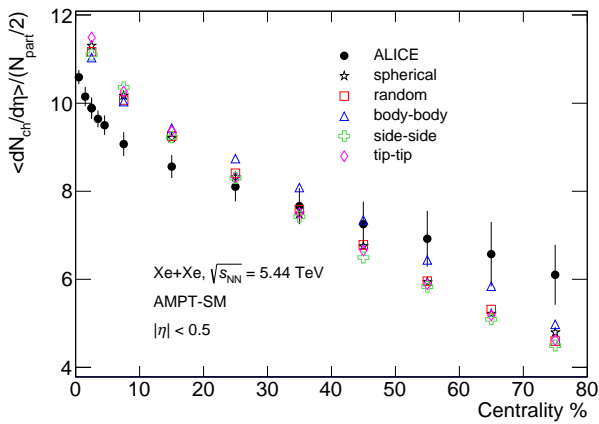}
\caption{(Color online) $<$dN$_{ch}$/d$\eta$$>$ per participating nucleons as a function of centrality for different Xe+Xe collision configurations from AMPT model. The model results are compared with the corresponding measurements from the ALICE experiment~\cite{alicemult}.}
\label{fig:dnchdetavsnpart} 
\end{figure}

\begin{figure}[hbtp]
\centering 
\includegraphics[scale=0.3]{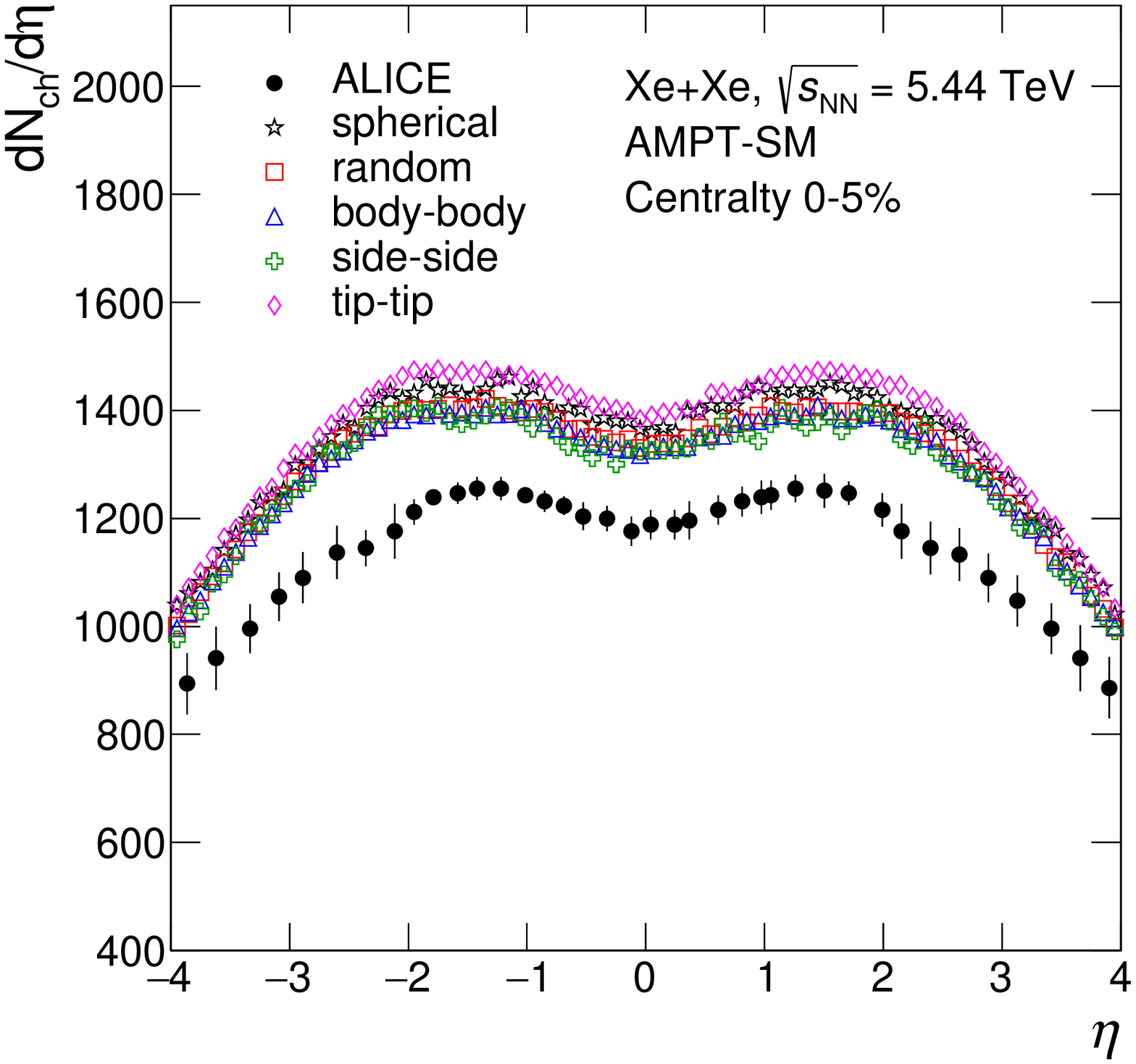}
\includegraphics[scale=0.3]{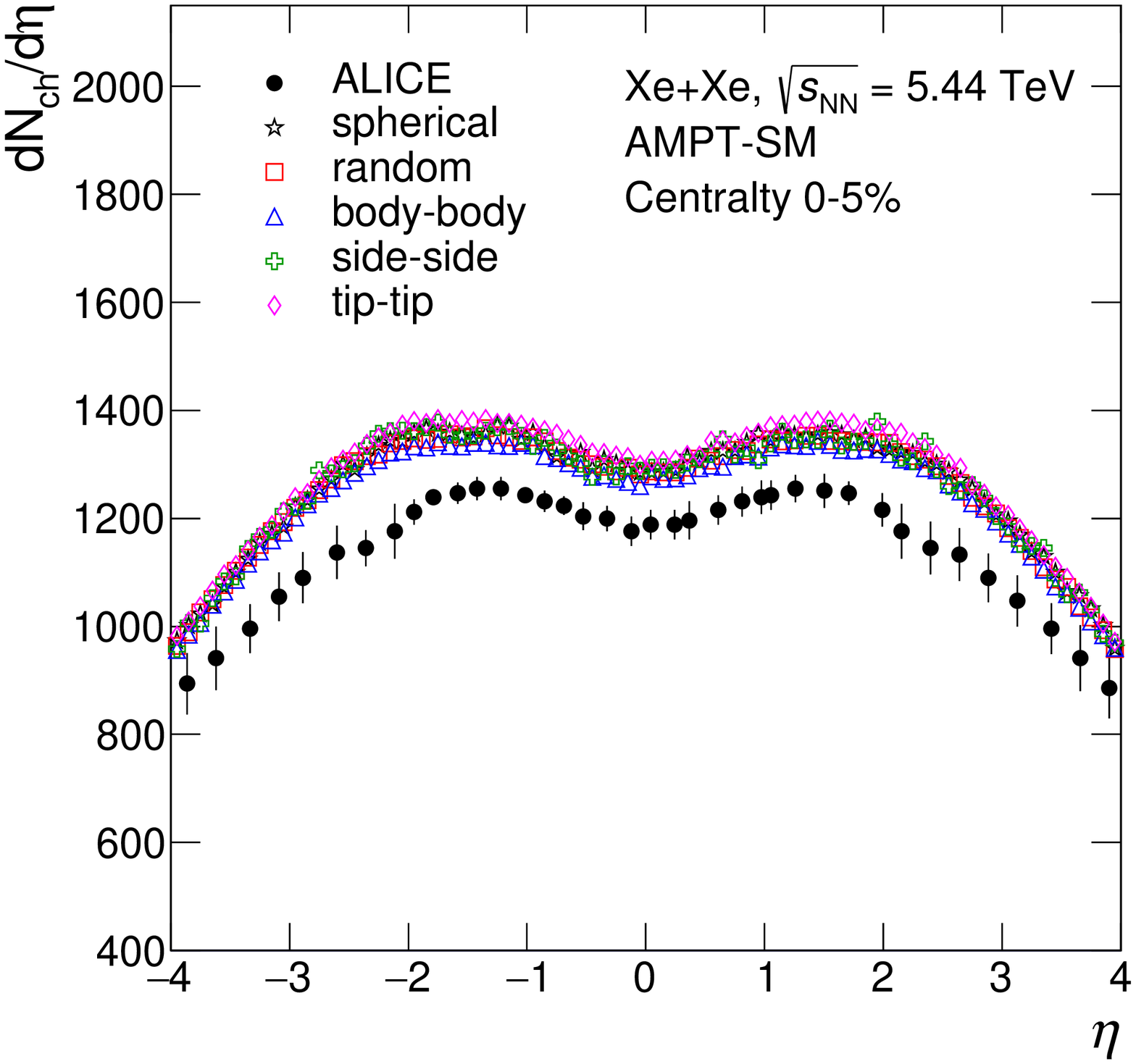}
\caption{(Color online) Upper panel: Pseudorapidity ($\eta$) distribution of inclusive charged particles in 0-5 $\%$ central Xe+Xe collisions at $\sqrt{s_{NN}}$ = 5.44 TeV for different configurations. Results from AMPT model are compared with the measurements from the ALICE experiment~\cite{alicemult}. Lower panel: Same as the upper panel but with centrality tuned in the model such that data and model have same $<$N$_{part}$$>$.}
\label{fig:eta} 
\end{figure}

\begin{figure*}[hbtp]
\centering 
\includegraphics[scale=0.7]{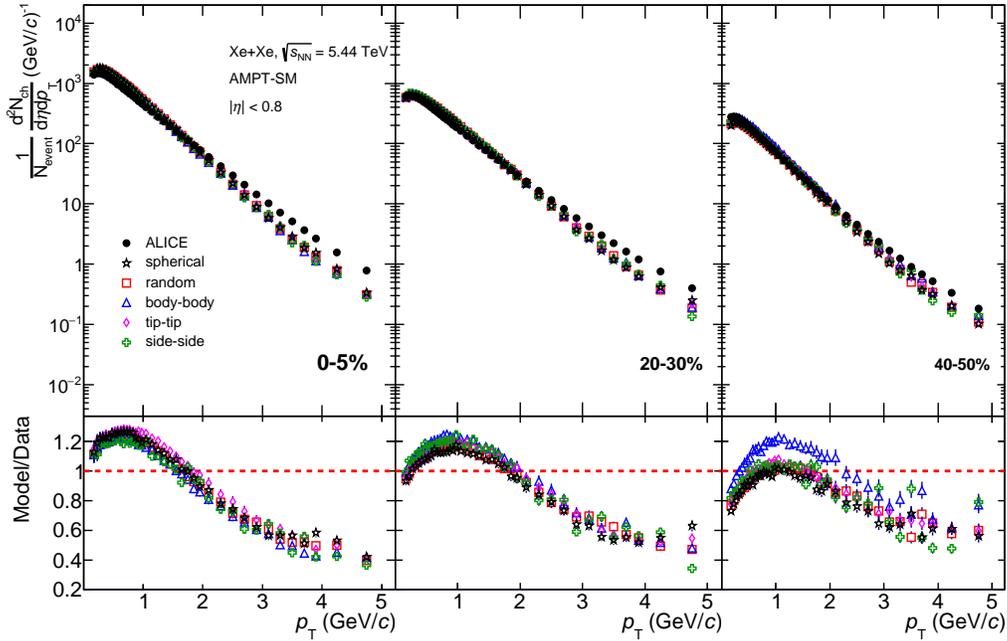}
\caption{(Color online) Upper panels show charged particle transverse momentum ($p_{\mathrm{T}}$) spectra at $|\eta|$ $<$ 0.8 in centrality class 0-5$\%$, 20-30$\%$ and 40-50$\%$ from AMPT model for various collision configurations in Xe+Xe collisions at $\sqrt{s_{NN}}$ = 5.44 TeV along with the measurement from the ALICE experiment~\cite{alicespectra}. Lower panels show the ratios of model to data.}
\label{fig:spectra} 
\end{figure*}

\begin{figure}[hbtp]
\centering 
\includegraphics[scale=0.4]{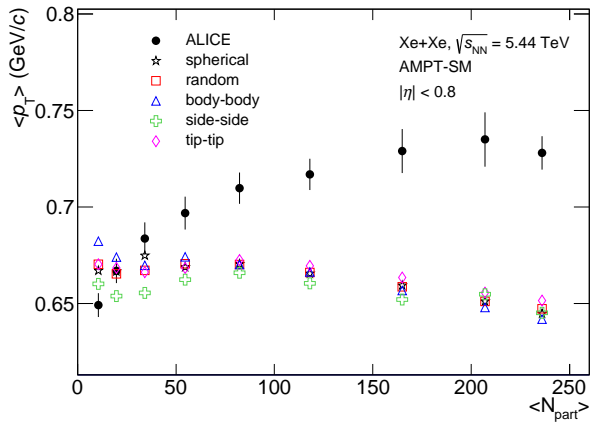}
\caption{(Color online) $<$$p_{\mathrm{T}}$$>$ of charged particle as a function of $<$N$_{part}$$>$ at $|\eta|$ $<$ 0.8 for various collision configurations in AMPT model along with the measurement from the ALICE experiment~\cite{alicespectra} in Xe+Xe collisions at $\sqrt{s_{NN}}$ = 5.44 TeV.}
\label{meanpt} 
\end{figure}

Figure~\ref{fig:dnchdetavsnpart} shows $<$dN$_{ch}$/d$\eta$$>$ per participating nucleons as a function of centrality in $|$$\eta$$|$ $<$ 0.5 for different types of collision configuration. Results from AMPT model are compared with the ALICE measurements~\cite{alicemult}. In central collisions, AMPT overpredicts the data and in peripheral collisions it underpredicts whereas for the mid-central collisions the results from AMPT have a good agreement with the data. $<$dN$_{ch}$/d$\eta$$>$/$(<N_{part}>/2)$ is similar in random and spherical case. Among the all collision configurations, the tip-tip configuration has largest charged particle multiplicity in central collisions whereas body-body configuration has largest charged particle multiplicity in mid central and peripheral collisions.\\
The upper panel of Fig.~\ref{fig:eta} shows $\eta$ distribution of produced charged particles in 0-5$\%$ central Xe+Xe collisions at $\sqrt{s_{NN}}$ = 5.44 TeV for various collision configurations whereas the lower panel shows the same but with a tuned centrality in the model so that the $<$N$_{part}$$>$ of data and model are the same. AMPT model is able to describe the shape of $\eta$ distribution but overpredicts the ALICE measurement~\cite{alicemult}.\\
Charged particle $p_{\mathrm{T}}$ spectra in the centrality class 0-5$\%$, 20-30$\%$ and 40-50$\%$ for Xe+Xe collisions at $\sqrt{s_{NN}}$ = 5.44 TeV are shown in Fig.~\ref{fig:spectra}. Results from AMPT model are compared with the measurements from the ALICE experiment~\cite{alicespectra}. For $p_{\mathrm{T}}$ $<$ 2.0 GeV/$c$ model overpredicts the data in central collisions, it is consistent with the data in mid central collisions and the model results underpredict the data for the peripheral collisions. For $p_{\mathrm{T}}$ $>$ 2.0 GeV/$c$ model underpredicts the data for all the centrality classes studied. The $p_{\mathrm{T}}$ spectra obtained from both spherical and random configurations are found to be consistent with each other.\\
Figure~\ref{meanpt} shows $<$$p_{\mathrm{T}}$$>$ of charged particles as a function of $<$N$_{part}$$>$ for various Xe+Xe collision configurations in AMPT model. We find that in AMPT model with string melting $<$$p_{\mathrm{T}}$$>$ does not show a strong centrality dependence and underpredicts the ALICE measurements except for the most peripheral collisions. In central and mid central Xe+Xe collisions the tip-tip configuration provides a higher value of $<$$p_{\mathrm{T}}$$>$ however in peripheral collisions body-body configuration leads to a higher $<$$p_{\mathrm{T}}$$>$ value compare to other configuration.
\subsubsection{\label{sec:level3}Elliptic and Triangular flow}
\begin{figure}[hbtp]
\centering 
\includegraphics[scale=0.4]{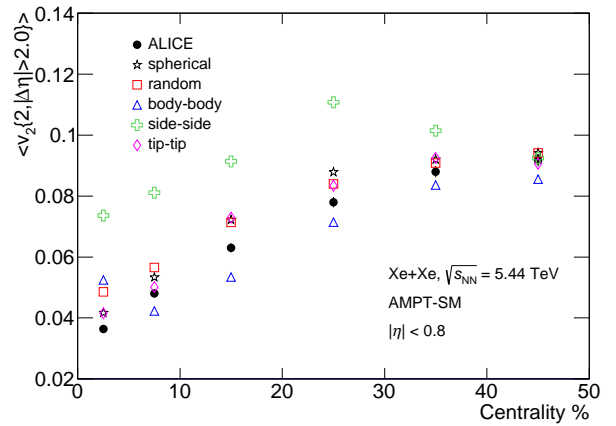}
\caption{(Color online) $p_{\mathrm{T}}$ integrated $<$$v_{2}$$>$ at $|\eta|$ $<$ 0.8 as a function of centrality from AMPT and ALICE experiment~\cite{xe} in Xe+Xe collisions at $\sqrt{s_{NN}}$ = 5.44 TeV.}
\label{v2cent} 
\end{figure}

\begin{figure}[hbtp]
\centering 
\includegraphics[scale=0.4]{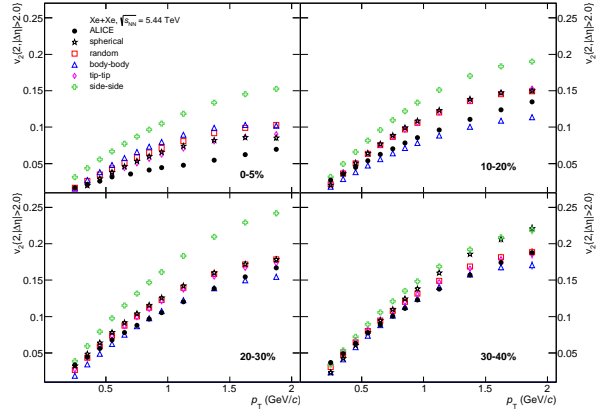}
\caption{(Color online) $p_{\mathrm{T}}$ differential $v_{2}$ at $|\eta|$ $<$ 0.8 from AMPT and ALICE experiment~\cite{xe} in Xe+Xe collisions at $\sqrt{s_{NN}}$ = 5.44 TeV}
\label{v2pt} 
\end{figure}

\begin{figure}[hbtp]
\centering 
\includegraphics[scale=0.4]{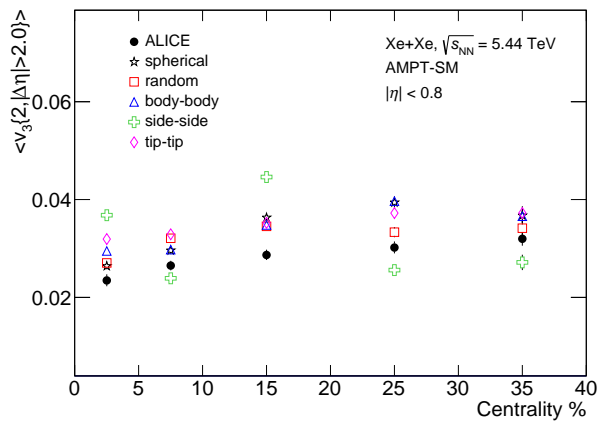}
\caption{(Color online) $p_{\mathrm{T}}$ integrated $<$$v_{3}$$>$ at $|\eta|$ $<$ 0.8 as a function of centrality from AMPT and ALICE experiment~\cite{xe} in Xe+Xe collisions at $\sqrt{s_{NN}}$ = 5.44 TeV.}
\label{v3cent} 
\end{figure}

Elliptic flow ($v_{2}$) and triangular flow ($v_{3}$) are respectively the second order and third order Fourier coefficients of particle distributions~\cite{v2gen}. We have calculated these flow coefficients by scalar product method. The n$^{th}$ order flow coefficient in scalar product method~\cite{scalar} is defined as,\\
\begin{equation}
 v_{n}\{2,|\Delta\eta|>2\}~=~\frac{<<u_{n,k}Q^{*}_{n}>>}{\sqrt{\frac{<Q_{n}Q^{A*}_{n}><Q_{n}Q^{B*}_{n}>}{<Q^{A}_{n}Q^{B*}_{n}>}}}
\end{equation}\\
where flow vector $Q_{n}$ = $\sum_{j=1}^{n} e^{in\phi_{j}}$ and $u_{n,k}$ = $e^{in\phi_{k}}$. $\phi_{j}$ is the azimuthal angle of j$^{th}$ particle and $k$ is the particle of interest. Double bracket ($<<$$>>$) corresponds to an average over all particles in all events and single bracket ($<$$>$) corresponds to an average over all events. $*$ is the complex conjugate. We calculate flow vectors $Q_{n}$, $Q^{A}_{n}$ and $Q^{B}_{n}$ in 2.8 $<$ $\eta$ $<$ 5.1, -3.7 $<$ $\eta$ $<$ 1.7 and $|\eta|$ $<$ 0.8 respectively. $|\Delta\eta|>2$ corresponds to the $\eta$ gaps between the regions where $u_{n,k}$, $Q_{n}$, $Q^{A}_{n}$ and $Q^{B}_{n}$ are calculated.\\
Figure~\ref{v2cent} shows $p_{\mathrm{T}}$ integrated $<$$v_{2}$$>$ for different Xe+Xe configurations in AMPT model along with the measurement from the ALICE experiment for Xe+Xe collisions at $\sqrt{s_{NN}}$ = 5.44 TeV. A clear centrality dependence of $<$$v_{2}$$>$ have been observed from AMPT model. $<$$v_{2}$$>$ in Xe+Xe collisions with randomly oriented Xe nucleus in AMPT model overpredicts the data~\cite{xe} in central collisions, however consistent with the ALICE measurements above 30$\%$ centrality. We find that $<$$v_{2}$$>$ in Xe+Xe collisions with randomly oriented deformed nucleus is 15$\%$ and 6$\%$ higher compared to the collisions with spherical Xe nucleus in 0-5$\%$ and 5-10$\%$ centrality and consistent with each other above collision centrality $10\%$. This observation is consistent with hydrodynamics prediction is given in~\cite{hydro}. side-side configuration leads to a larger $<$$v_{2}$$>$ compare to all other configurations. Tip-tip configuration shows smallest $<$$v_{2}$$>$ in 0-5$\%$ central collisions whereas for other centrality studied the body-body configuration gives lowest $<$$v_{2}$$>$. We have also studied $p_{\mathrm{T}}$ differential $v_{2}$ in different centrality classes and those are shown in Fig~\ref{v2pt}. A strong $p_{\mathrm{T}}$ dependence of $v_{2}$ has been observed.
Figure~\ref{v3cent} shows $p_{\mathrm{T}}$ integrated $<$$v_{3}$$>$, for 5 different Xe+Xe collision configurations in AMPT model  along with the measurements from the ALICE experiment. A mild centrality dependence in measured  $<$$v_{3}$$>$ is observed from AMPT model. AMPT model with random configuration overpredicts the ALICE measurements~\cite{xe}.
\section{\label{sec:level4}Summary}
In this work, we have implemented the deformation of Xe nucleus in AMPT model framework which gives us a unique possibility to study the different type of collision configuration. These collision configurations are sensitive to initial conditions and can be used to constrain various initial state models. Deformation of Xe nucleus is implemented by using a deformed Woods-saxon profile of nucleon distribution. We have studied 3 special type of collision configurations of Xe nucleus which have been achieved by rotating the Xe nucleus in polar and azimuthal directions. We compare various result from AMPT model with the corresponding ALICE measurements. To understand the effect of deformation on the measured observable, we compare the AMPT model results from the collisions of deformed Xe nucleus with the results from the collisions where Xe nucleus is conder to be spherically symmetric. In addition to that we also compare the results between different Xe+Xe collision configurations.\\
We find that the observed $<$dN$_{ch}$/d$\eta$$>$ at $|\eta|$ $<$ 0.5 in AMPT model overpredicts the ALICE measurements in central Xe+Xe collisions at $\sqrt{s_{NN}}$ = 5.44 TeV, consistent with the data in mid central collisions and underpredicts the ALICE measurements in peripheral collisions. The calculated $<p_{T}>$ from AMPT model does not show any significant centrality dependence and underpredicts the ALICE measurements for all centrality classes except for the most peripheral collisions. random configuration of AMPT model gives a slightly higher value of $v_{2}$ compare to ALICE measurements in central collisions, however it is consistent with the data above 30$\%$ centrality. Measured $v_{3}$ in ALICE experiment is overpredicted by the results from the random Xe+Xe configuration of AMPT model.\\
We observe a dependence on the Xe+Xe collision configurations for the measured observables. In central collision the tip-tip configuration leads to a higher value of $<$dN$_{ch}$/d$\eta$$>$ compared to other configurations, whereas in peripheral collision the body-body configuration gives the largest value of $<$dN$_{ch}$/d$\eta$$>$. In central and mid central Xe+Xe collisions the tip-tip configuration provides a higher value of $<$$p_{\mathrm{T}}$$>$ however in peripheral collisions the body-body configuration leads to a higher $<$$p_{\mathrm{T}}$$>$ value compare to other configurations.We do not observe any significant change in charged particle multiplicity and $<$$p_{\mathrm{T}}$$>$ between the spherical case and random case. Elliptic flow in the central collision is enhanced by $\sim$15$\%$ for random configuration compared to spherical configuration.  In addition to that we also find that the side-side configuration of Xe+Xe collisions lead to the largest value of $v_{2}$ and body-body configuration gives a smaller value of $v_{2}$ relative to other configurations. The calculated elliptic flow in side-side configuration is more than 50$\%$ larger compared to other configuration in central and mid-central collisions.\\
If ALICE experiment could trigger these different collision configurations then these different initial state condition can be possibly accessed. The future scope of our study includes the study of CME observable in central body-tip type collisions where we expect a finite amount of magnetic field and less value of $v_{2}$. We would also like to develop a way to select such special types of collision configurations in the experiment.
\section{\label{sec:level5}Acknowledgwments}
We would like to acknowledge Dr. Rihan Haque for helping to implement the deformation of Xe nucleus in AMPT model and Dr. Md. Nasim for valuable discussion. We also acknowledge Dr. Z. W. Lin for help in the AMPT simulations.


\begin{thebibliography}{50}
\bibitem{v2} J. Y. Ollitrault, Phys. Rev. D 46 (1992) 229.
\bibitem{jet} X. N. Wang, Nucl. Phys. A 98 (2005) 750.
\bibitem{rhic} J. Adams et al., [STAR Collaboration], Nucl. Phys. A 757 (2005) 102.
\bibitem{alice} U. Heinz and M. Jacob, arXiv:0002042 [nucl-th].
\bibitem{Pb} K. Aamodt et al., [ALICE Collaboration], Phys. Rev. Lett. 105 (2010) 252301.
\bibitem{xe} S. Acharya et al., [ALICE Collaboration], Phys. Lett. B 784 (2018) 82.
\bibitem{deform} P. M ̈oller, A. J. Sierk, T. Ichikawa and H. Sagawa, arXiv:1508.06294 [nucl-th].
\bibitem{hydro} G. Giacalone, J. N. Hostler, M. Luzum and J. Y. Ollitrault, Phys. Rev. C 97 (2018) 034904.
\bibitem{cme1} D. Kharzeev, Phys. Lett. B633 (2006) 206.
\bibitem{cme2} D. E. Kharzeev, L. D. McLerran and H. J. Warringa, Nucl. Phys. A 803 (2008) 227.
\bibitem{alicespectra} S. Acharya et al., [ALICE Collaboration], arXiv:1805.01832 [nucl-ex].
\bibitem{ampt1} Z. W. Lin, C. M. Ko, B. A. Li, B. Zhang, and S. Pal, Phys. Rev. C 72 (2005) 064901.
\bibitem{ampt2} J. Xu and C. M. Ko, Phys. Rev. C 83 (2011) 034904.
\bibitem{hijing} X. N. Wang and M. Gyulassy, Phys. Rev. D 44 (1991) 350.
\bibitem{zhang} B. Zhang, Comput. Phys. Commun. 109 (1998) 193.
\bibitem{art} B. A. Li and C. M. Ko, Phys. Rev. C 52 (1995) 2037.
\bibitem{wood} K. Hagino, N. W. Lwin, and M. Yamagami, Phys. Rev. C  74 (2006) 017310.
\bibitem{uurihan} Md. R. Haque, Z. W. Lin, B. Mohanty, Phys. Rev. C 85 (2012) 034905.
\bibitem{alicemult} S. Acharya et al., [ALICE Collaboration], arXiv:1805.04432 [nucl-ex].
\bibitem{v2gen} S. Voloshin and Y. Zhang, Z. Phys. C 70 (1996) 665.
\bibitem{scalar} C. Adler et al., [STAR Collaboration], Phys. Rev. C 66 (2002) 034904.
\end{thebibliography}
\end{document}